\documentstyle[11pt,aasms4]{article}
\begin{document}
\title{Differences between the Two Anomalous X-Ray Pulsars: 
Variations in the Spin Down Rate of 1E 1048.1-5937 and
An Extended Interval of  
Quiet Spin Down in 1E 2259+586  
 }

   \author{Altan Baykal $^{1}$, 
          Tod Strohmayer $^{2}$,
          Jean Swank $^{2}$,
          M. Ali Alpar $^{3}$,
          Michael J. Stark $^{4}$ }  
\affil{ 
 $^{1}$ Physics Department, Middle East Technical University,
  Ankara 06531, Turkey\\ 
 $^{2}$ Laboratory for High Energy Astrophysics NASA/GSFC
  Greenbelt, Maryland 20771 USA \\
 $^{3}$ Department of Physics, Sabanc{\i} University, Istanbul, Turkey \\
 $^{4}$ Department of Physics, Marietta College, Marietta, OH  45750, USA  }

   \begin{abstract}

 We analysed the RXTE archival data of 1E 1048.1-5937
covering a time span of more than one year. The spin down rate
of this source decreases by $\sim$30$\%$ during the observation.
We could not resolve the X-ray
flux variations because of contamination by Eta Carinae.
We find that the level of pulse frequency fluctuations 
of 1E 1048.1-5937 is consistent with typical noise levels of accretion
 powered pulsars (Baykal $\&$ {\"O}gelman 1993, Bildsten et al., 1997).
Recent RXTE observations of 1E 2259+586 have shown  
a constant spin down
with a very low upper limit on timing noise (Kaspi et al., 1999).
We used the RXTE archival X-ray observations of 1E 2259+586
to show that
the intrinsic X-ray luminosity times series  
is also  stable,
with an rms fractional variation of less than 15$\%$.
The source could have been in a quiet phase of accretion 
with a constant
X-ray luminosity and spin down
rate.


\end{abstract} 
\section{Introduction}
The two sources considered in this paper belong to a small 
group of X-ray pulsars which are called 
anomalous X-ray pulsars (hereafter AXPs), with periods in the
 $\sim$5--12 s range. 
This narrow pulse period distribution of the AXPs, is significantly different
from that of High Mass X-Ray Binaries (HMXRBs) where  
the pulse periods span a range from 69 ms to 25 min.   
Their X-ray spectra have   
steep power law indices $\sim$ 3-4 in addition to soft blackbody components 
with kT$\sim$0.5 keV in some of the sources (Stella et al., 1998).
They lack observed optical 
counterparts. Their X-ray luminosities 
are at the order of 10$^{35}$-10$^{36}$ erg s$^{-1}$ and 
spin down rates are relatively constant. 

The first proposed model for AXPs was accretion from
low-mass X-ray companions at lower accretion rates
(L$_{x}=GM\dot M/R \sim 1\times 10^{35}$erg s$^{-1}$) with magnetic
fields of B$\sim 10^{11}$ Gauss (Mereghetti $\&$ Stella 1995).
In this scenario, the observed pulse periods of AXPs 
can be explained as rotation 
periods of neutron stars close to the equilibrium periods 
for accretion from a disk.
However,  orbital signatures such as periodic delays in pulse 
arrival times or periodic flux changes have not been observed
in the AXPs. 
This has led several researchers to 
alternative interpretations based on the single pulsar 
hypothesis. Corbet et al. (1995) suggested the 
possibility of accretion from a molecular cloud.
Alternatively, AXPs could be isolated stars which are
accreting from a disk formed
as remnants of the common envelope evolution
of HMXRBs (van Paradijs et al. 1995, Ghosh et al. 1997).
On the other hand 
Thompson $\&$ Duncan (1993) proposed that these sources 
are highly magnetized
 ($\sim$10$^{14}$-$10^{15}$ Gauss) isolated neutron stars (magnetars)
which are slowing down due to electromagnetic dipole radiation. 
 According to the magnetar theory there should be  
several unseen very large glitches in  
pulse period histories of 1E 1048.1-5937 and 1E 2259+586
 (Heyl $\&$ Hernquist 1999). 
These glitches should be at least a 
factor of hundred larger than the radio pulsar glitches 
in $\delta \nu /\nu $ 
(Thompson $\&$ Duncan 1996, Heyl $\&$ Hernquist 1999). 
     
1E 1048.1-5937 was discovered by the Einstein 
satellite during the observations of the Carina nebula 
(Seward, Charles $\&$ Smale 1986).
1E 2259+586 is located at the center
of the radio/X-ray supernova remnant G109.1-1.0 (Fahlman \& Gregory
1981).
 Both sources lack bright optical counterparts
 (Mereghetti, Caraveo $\&$ Bignami 1992, Coe $\&$ Jones 1992).
 If they are binary systems, 
their companion stars should be either 
white dwarfs or helium-burning stars with M$<$0.8M$_{o}$ 
(Mereghetti, Israel, Stella, 1998, Baykal et al. 1998).  

The torque changes of 1E 1048.1-5937 and 1E 2259+586 were studied by 
Mereghetti (1995), Corbet $\&$ Mihara (1997) and Baykal $\&$ Swank (1996). 
Both sources showed pulse frequency changes which can support 
the accretion hypothesis.
In this work,
we present two new pulse frequency measurements from
long observations in the  archival RXTE data base.
In the $\sim$400 day  time span of the observation, 
we found the spin-down rate of 1E~1048.1--5937
to change by 30 \%.
Recent pulse timing analysis of 1E~2259+586 
(Kaspi et al., 1999) has shown that the source had constant 
spin down over a 2.6 yr time span, with very low
timing noise. 
We extracted archival data 
of 1E 2259+586 and constructed a bolometric X-ray luminosity time series.
We found that 
the X-ray luminosity is almost constant while the spin-down rate 
is constant (Kaspi et al., 1999). 

\section{Data Analysis} 

The archival observations of 1E 1048.1-5937 and 1E 2259+586  
are listed in Table 1.
The results presented here 
are based on data collected with the Proportional Counter Array 
(PCA, Jahoda et al., 1996). The PCA instrument consists of an array 
of 5 proportional counters operating in the 2-60 keV energy range, with 
a total effective area of approximately 7000 cm$^{2}$ and a field of view 
of
$\sim 1^{\circ}$ FWHM.

Background light curves and the pulse height amplitudes were generated using
the background estimator models based on the rate of very large events (VLE),
spacecraft activation and cosmic X-ray emission with the standard PCA
analysis tools (ftools) 
and were subtracted from the source light curve obtained from
the first Good Xenon layer of event data. The background subtracted light
curves were corrected with respect to the barycenter of the solar system. 
From the long archival data string,   
pulse periods for  
1E 1048.1-5937 were found by folding the time series 
on statistically independent trial periods (Leahy et al. 1983).
Master pulses were constructed
from these observations
by folding the data on
the period giving the maximum $\chi^2$.
The master pulses were arranged in 55 phase
bins and represented by their Fourier harmonics (Deeter \& Boynton 1985) and
cross-correlated with the harmonic representation of average pulse profiles
from each observation. The pulse arrival times are obtained from the 
cross-correlation analysis. 
 The linear trend of pulse arrival times is a direct
measure of the pulse frequency during the observation, 
\begin{equation}
\delta \phi = \phi_{o} + \delta \nu (t-t_{o})
\end{equation}
where $\delta \phi $ is the pulse phase offset deduced from the pulse
timing analysis, $t_{o}$ is the mid-time of the observation, $\phi_{o}$ is
the phase offset at t$_{o}$, $\delta \nu$ is the deviation from the mean
pulse frequency (or additive correction to the pulse frequency). 
The pulse period measurments of 1E 1048.1-5937 from archival RXTE observations
are presented in Table 2.

 During the RXTE observations the pulse frequency derivative
of 1E 1048.1-5937 decreased approximately by 30 \% (see Table 2).
 However these changes
cannot be correlated
 with X-ray flux changes since Eta Carinae lies 45' a way from
1E 1048.1-5937. The  strong flux changes of
Eta Carinae (Corcoran et al., 1997) contaminate the FOV of
 1E 1048.1-5937 since the FWHM of RXTE/PCA $\sim 1^{\circ}$. It should
also be pointed out
that in the archival RXTE
observations, there are found short,
$\sim$ 2000 sec, observations separated by months;
due to the pulse frequency derivative changes we have not
phase connected them in order to avoid any cycle count
ambiguity. Similarly, two successive Ginga observations
of 1E 1048.1-5937 which were separated by 10 days cannot be combined
in phase because of the cycle count ambiguity (Corbet $\&$ Day 1990).

Einstein, EXOSAT and Ginga observations of 1E 1048.1--5937 provided 5
pulse frequency measurements over 10 years, which were consistent with
a constant spin-down rate of  $\dot \nu \sim 3.8 \times 10^{-13}$
Hz s$^{-1}$ (Mereghetti 1995).
This is
$\sim $38  times higher than that of 1E 2259+586 (Baykal et al., 1998).
 Observations with ROSAT
in 1992-1993 indicated that the spin down rate almost doubled
from its value in 1988
(Mereghetti 1995). The mean flux decreased by a factor of
3 compared to the value measured with EXOSAT
in mid of $\sim$ 1985 (Corbet $\&$ Mihara 1997).
These variations are consistent with
accretion-powered X-ray emission.

In order to deduce the pulse frequency changes
 of 1E 1048.1-5937, the residuals of pulse frequencies
are extracted from their linear trends and the residuals
are presented in terms of their sigma values.
Fig. 1 (lower) 
clearly shows that
 the pulse frequency fluctuations are significant at the order of several
 $\sigma $ levels.
The residual pulse frequencies
 between $\sim$ 48600 -- MJD $\sim$ 50800 MJD yield a noise strength
on the  order of
$ S\approx (2\pi)^{2}<\delta \nu ^{2}>/T 
\approx (2\pi)^{2}<\delta \phi ^{2}>/T^{3} 
 \sim 10^{-17}$ rad$^{2}$ sec$^{-3}$, 
where  $<\delta \nu ^{2}>$ and $<\delta \phi ^{2}>$ 
are the normalized
variances of residual pulse frequencies and pulse arrival times and 
$T$ is the total time span 
(see Cordes 1980 for further definitions of noise strength).
 This value is
comparable with typical
accretion powered pulsar noise strength Baykal $\&$ {\"O}gelman (1993).

Recent RXTE observations of 1E 2259+586 have shown 
a constant spin down rate
with a low  upper limit on timing noise (Kaspi et al., 1999). 
The residuals of the pulse arrival times gives an 
upper limit to the noise strength at 
$T_{observation} \sim 2.6$ years, 
S$\approx (2\pi)^{2}<\delta \phi ^{2}>/T^{3}_{observation}
\sim 10^{-24}$ rad$^{2}$ sec$^{-3}$. 
 This value is 5 decades lower than 
the value which is deduced from
15 years of pulse period history of 1E 2259+586
 (Baykal $\&$ Swank 1996, Baykal $\&$ {\"O}gelman 1993)
 and 2 orders lower 
than that of the Crab pulsar (Boynton et al., 1972)
or that of the 
LMXRB pulsar 4U 1626-67 (Chakrabarty 1997). This upper limit 
is indeed very low  
for an accretion-powered X-ray pulsar. 
If the pulse period changes are due to variations in the  
accretion process, the X-ray luminosity 
would be a constant (Ghosh $\&$ Lamb 1979). In order to check the 
variations in its X-ray luminosity, we derived the X-ray luminosities for 
all observations from the X-ray spectra. 

The X-ray background spectrum was calculated using the 
 background estimator
models based upon the rate of very large events (VLE), spacecraft
activation and cosmic X-ray emission as used to calculate background
light curves. The X-ray spectra are fitted with a power law spectrum with 
a photon index 4.78 and
column density
 N$_{H}$=(2.2$\pm$ 0.8)$\times$ 10$^{22}$ cm$^{-2}$, parameters  
 consistent with those obtained 
from ASCA and SAX measurements in the 2-10 keV range
(Corbet et al., 1995, Parmar et al., 1998).
 The resultant background subtracted X-ray 
luminosity time series in the energy 
range 2-10 keV is presented in Fig. 2. The variation in the bolometric X-ray 
luminosity is less than $\%15$. This is   
consistent with low timing noise and secular spin down
 according to accretion models.   

In Ginga observations 1E~2259+586 had flux levels a factor of two higher
than average (Iwasawa et al. 1992).  This implied that, if due to accretion,
the 
rate onto the source was variable and fluctuations in the spin down
rate would be expected.
Indeed for this era high noise levels $S\sim 10^{-19}$rad$^{2}$sec$^{-3}$ 
are found
 (Baykal $\&$ Swank
1996).

\section{Discussion}

 In the pulse timing analysis of 1E 1048.1-5937,
we found
pulse frequency changes on the  time scale of  400 days.
The source spin down rate has doubled since 1992 
(Mereghetti 1995, Corbet $\&$ Mihara 1997).
The pulse frequency measurements since 1992
show significant fluctuations from the average spin down trend.
The level of pulse frequency
fluctuations of 1E 1048.1-5937 are found to be consistent with 
typical noise levels of accretion
powered pulsars (Baykal $\&$ {\"O}gelman 1993).

Two Soft Gamma ray Repeaters (SGR)
SGR 1806-20, SGR 1900+14 were recently identified as magnetars 
(Kouveliotou et al. 1998; 1999).
The similarity of SGR and AXP
 pulse periods and secular spin down trends
raised the question 
whether all AXPs are actually magnetars.
According to the magnetar theory   
pulse frequency fluctuations may be caused by processes 
such as: variable emission of Alfven waves or particles from magnetars 
(Thompson $\&$ Blaes 1998), radiative precession of magnetars 
(Melatos 1999), or discontinous spin up events (glitches) like those
seen 
in radio pulsars (Thompson $\&$ Duncan 1996, see also Kaspi et al., 2000). 
Heyl $\&$ Hernquist (1999) investigated the pulse frequency 
histories of 1E 1048.1-5937 and 1E 2259+586 and proposed that 
the irregularities of these sources are explained by 
several large glitches  
with sizes $\delta \nu  
\sim 2 \times 10^{-4}$ s$^{-1}$ and $\sim 5 \times 10^{-6}$ s$^{-1}$,
respectively.
While the values of $\delta \nu$ are only about 20 times those of 
the Vela and Crab pulsars, respectively, the  
values of $\delta \nu /\nu$ 
are several orders of magnitude larger than radio pulsar glitches and 
they occur much more frequently than scaling 
from the statistics of large  
radio pulsar glitches would lead us to expect
(Alpar $\&$ Baykal 1994). 

In summary, both sources have shown pulse frequency fluctuations on the
order of a few decades (Mereghetti 1995, Baykal $\&$ Swank 1996).
The level of torque changes of 1E 2259+586 over a time scale
of 15 years is consistent with
accretion-powered X-ray binaries (Baykal $\&$ Swank 1996).
 The steady level of the bolometric X-ray luminosity during the 
quiet spin down epoch (Kaspi et al., 1999)
 is consistent with accretion models.
 Alternatively 
it may turn out to be made out of discrete glitches over
 a background of relatively quiet spin down, 
as in isolated pulsars and magnetars (Heyl $\&$ Hernquist 1999).
 Continued phase coherent X-ray timing of this source should 
prove extremely important in finally deciding its nature.
The correlation of the spin-down rate of
1E 1048.1-5937 
with the X-ray flux supports
the possibility that it is an accreting source
(Mereghetti 1995, Corbet $\&$ Mihara 1997). Our work indicates that 
the source has high timing noise.
To see the exact nature of correlations between X-ray flux
and pulse frequency derivatives, an even more
extensive broad band X-ray observation should be carried out.

{\bf Acknowledgements~:}\\
We thank the referee for helpful comments.

{\Large{\bf Figure Caption}}\\

{\bf Fig. 1}~Pulse frequency time series of 1E 1048.1-5937. The solid
lines (upper panel) are fits to the measurements before and after Sep
1988, respectively. Residuals from these linear trends are shown in
terms of sigma values (lower panel). Measurements with Einstein,
EXOSAT, and Ginga (x) fit a trend well, while the later measurements
with ROSAT and ASCA (open circles) and RXTE (filled circles) fit a
trend of higher spin-down, but with well measured deviations.

{\bf Fig. 2}~X-ray luminosity time series of 1E 2259+586.

\begin{table}
\caption{RXTE observations of 1E 1048.1-5937 and 1E 2259+586}
\label{Pri}
\[
\begin{tabular}{ c c c }  \hline \hline
Time of Observation & Exposure      \\
 mm/dd/yy           &  sec            \\ \hline \hline
Source Name         & 1E 1048.1-5937  \\ \hline \hline
29/07/96-31/07/96   & 62586             \\
08/03/97            & 21165            \\
08/10/97-09/10/97   & 14971             \\ \hline \hline
Source Name         & 1E 2259+586     \\ \hline \hline
29/09/96-01/10/96   & 74758         \\
25/12/96            & 928            \\
25/01/97            & 1004            \\
22/02/97            & 1076             \\
25/02/97-26/03/97   & 100237            \\
18/04/97            & 727                \\
10/05/97            & 988                 \\
18/06/97            & 922                  \\
17/07/97            & 847                   \\
12/08/97            & 729                    \\
19/09/97            & 1038                    \\
16/10/97            & 832                      \\
14/11/97            & 884                       \\
13/08/98-02/12/98   & 121833                     \\ \hline \hline
\end{tabular}
\]
\end{table}

\begin{table}
\caption{RXTE Pulse Period Measurements of 1E 1048.1-5937
   }
\label{Pri}
\[
\begin{tabular}{ c c c }  \hline
Epoch(MJD)  & Pulse Period (sec)   &      \\ \hline
50294.67    & 6.449769$\pm$0.000004& Mereghetti et al., (1998)  \\
50515.69    & 6.450198$\pm$0.000018& This work   \\
50729.71    & 6.450486$\pm$0.000018& This work  \\ \hline
Time Span (MJD)   & Derivative of Pulse Frequency Hz sec$^{-1}$ \\ \hline
50294.67-50729.71 &$ -(4.74 \pm 0.18) \times 10^{-13} $       \\
50294.67-50515.69 &$ -(5.40 \pm 0.38) \times 10^{-13} $       \\
50515.69-50729.71 &$ -(3.72 \pm 0.52) \times 10^{-13} $        \\ \hline

\end{tabular}
\]
\end{table}

\newpage
\clearpage
\begin{figure}
\plotone{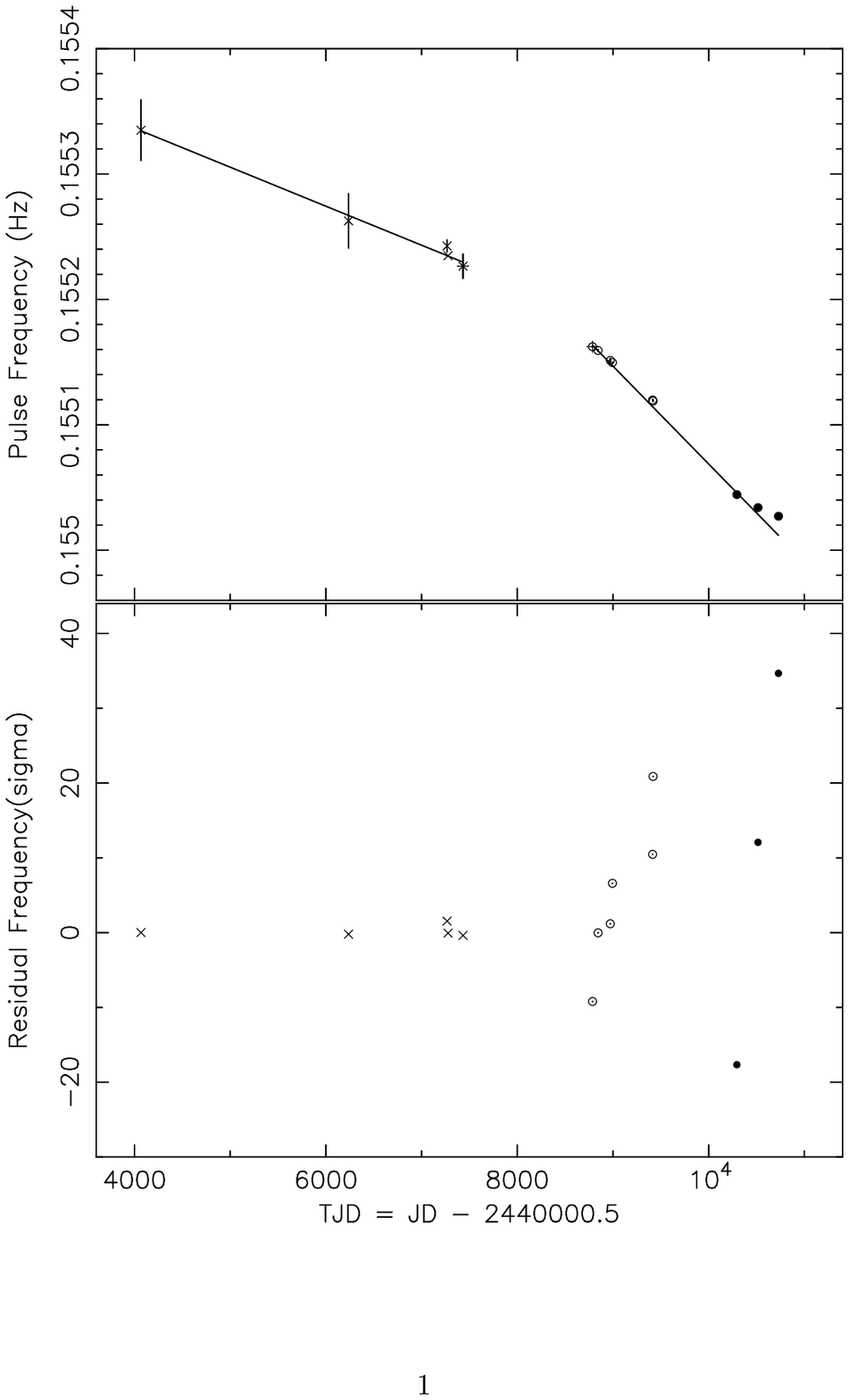}
\end{figure}

\newpage
\clearpage
\begin{figure}
\plotone{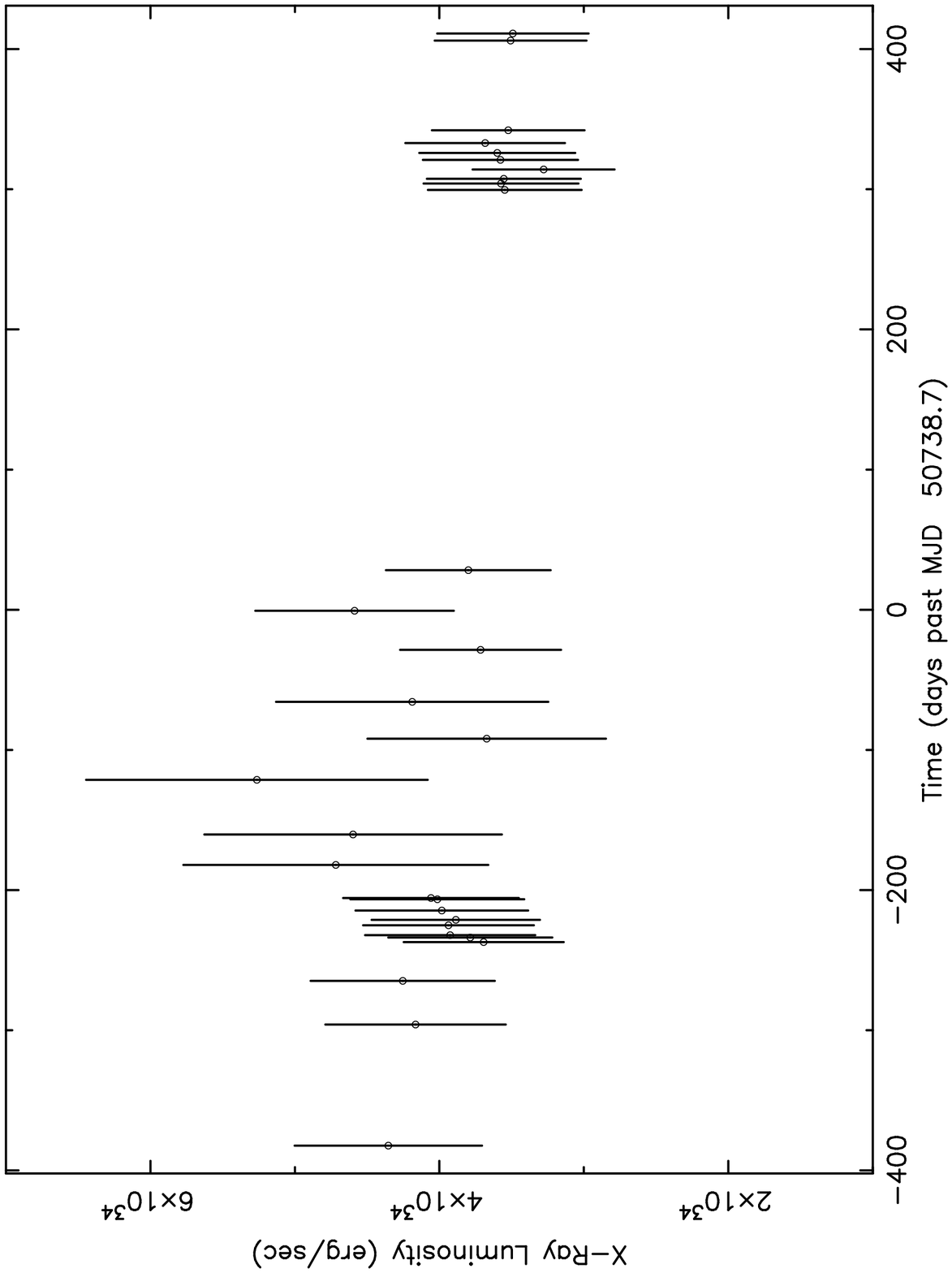}
\end{figure}


\begin{thebibliography}{}

\bibitem{} Alpar M.A., Baykal  A., 1994, MNRAS, 269, 849

\bibitem{} Baykal  A., {\"O}gelman, H, 1993,
                A\&A, 267, 119

\bibitem{} Baykal  A., Swank  J., 1996, ApJ, 460, 470

\bibitem{} Baykal  A., Swank J.H., Strohmayer T., Stark M.J., 
           1998, 336, 173

\bibitem{} Bildsten L., et al., 1997, ApJS, 113, 367

\bibitem{} Boynton  P.E., et al,  
           ApJ, 1972, 175, 217 

\bibitem{} Chakrabarty  D., et al.,
                1997, ApJ, 474, 414 

\bibitem{} Corbet R.H.D., Day C.S.R., 1990, MNRAS, 259, 191

\bibitem{} Cordes J.M., 1980, ApJ, 237, 216

\bibitem{} Coe M.J., Jones L.R., 1992, MNRAS, 259, 191

\bibitem{} Corbet  R.H.D., Smale A.P., Ozaki M.,
                Koyama K., Iwasawa K., 1995, ApJ,
                443, 786 

\bibitem{} Corbet  R.H.D., Mihara T., 1997, ApJ, 475, L127

\bibitem{} Corcoran, M.F., et al., 1997, Nature, 
           390, 587

\bibitem{} Deeter  J.E., Boynton  P.E., 1985,
                in Proc. Inuyama Workshop on Timing
                Studies of X-Ray Sources, ed. S. Hayakawa
                $\&$ F. Nagase (Nagoya: Nagoya Univ.), 29 

\bibitem{} Fahlman G.G., Gregory P.C., 1981, Nat, 293, 202

\bibitem{} Ghosh P., Angelini L., White N.E.,
                1997, ApJ, 478, 713 

\bibitem{} Ghosh P., Lamb F.K., 1979, ApJ., 232, 259

\bibitem{} Jahoda, K., Swank, J., Giles, A.B.,
                Stark, M.J., Strohmayer, T., Zhang, W.,
                1996, Proc. SPIE, 2808, 59

\bibitem{} Heyl, S.H., Hernquist L., 1999, MNRAS, 304, L37

\bibitem{} Iwasawa K., Koyama, K., Halpern, J.P.
           1992, PASJ, 44, 9

\bibitem{} Kaspi V.M, Chakrabarty D., Steinberger J., 
           1999 ApJL 525, L33

\bibitem{} Kaspi V.M, Lackey J.R, Chakrabarty, D, 2000,
            astro-ph/0005326, accepted for ApJ Letters 

\bibitem{} Kouveliotou C., et al., 1998, Nature, 393, 325

\bibitem{} Kouveliotou C., et al., 1999, ApJ, 510, L115

\bibitem{} Melatos A., 1999, ApJ., 519, L77 

\bibitem{} Mereghetti S., Caraveo P., Bignami G.F., 1992, A$\&$A, 263, 172

\bibitem{} Mereghetti S., 1995, ApJ, 455, 598 

\bibitem{} Mereghetti S., Stella L., 1995, ApJ, 442, L17

\bibitem{} Mereghetti S., Israel G.L., Stella L., 1998, MNRAS, 269, 689

\bibitem{} Leahy, D.A., et al.,
                1983, ApJ.,
                266, 160

\bibitem{} Parmar, A.N., et al., 
           1998, A\&A, 330, 175 

\bibitem{} Seward F., Charles P.A., Smale A.P., 1986, ApJ, 305, 814 

\bibitem{} Stella L., Israel G.L., Mereghetti S., 1998, Adv. Space Res, 
          vol 22, No. 7, pp 1025 

\bibitem{} Thompson C., Blaes O., 1998, Phys. Rev D, 57, 3219

\bibitem{} Thompson C., Duncan R.C., 1993, ApJ, 408, 194 

\bibitem{} Thompson C., Duncan R.C., 1996, ApJ, 473, 322

\bibitem{}  van Paradijs., J, Taam, R.E., van den Heuvel, E.P.J.,
                1995, A\&A, 299, L41


\end{thebibliography}
\end{document}